\documentclass[journal]{IEEEtran}
\IEEEoverridecommandlockouts

\usepackage{amsmath}
\usepackage{amsfonts}
\usepackage{amssymb}
\usepackage{graphicx}
\usepackage{epsfig}
\usepackage{float}
\usepackage{epstopdf}
\usepackage{array}
\usepackage{multirow}
\usepackage{setspace}
\usepackage{color}
\usepackage[section]{placeins}
\usepackage{url}
\usepackage{algorithm}
\usepackage{algpseudocode}
\usepackage{multicol}
\usepackage{soul}
\usepackage{comment}
\usepackage[normalem]{ulem}
\usepackage{tikzpagenodes}

\title{Fronthaul Compression Control for shared Fronthaul Access Networks
 }
\author{
  \IEEEauthorblockN{
    ~Sandra~Lag\'en\IEEEauthorrefmark{1},~Xavier~Gelabert\IEEEauthorrefmark{2},~Andreas~Hansson\IEEEauthorrefmark{2},~Manuel~Requena\IEEEauthorrefmark{1},~Lorenza~Giupponi\IEEEauthorrefmark{1}}\\
    \IEEEauthorblockA{\IEEEauthorrefmark{1}Centre Tecnol\`ogic de Telecomunicacions de Catalunya (CTTC/CERCA), Barcelona, Spain\\}
	\IEEEauthorblockA{\IEEEauthorrefmark{2}Huawei Technologies Sweden AB, Kista, Sweden \\
	\{sandra.lagen, manuel.requena, lorenza.giupponi\}@cttc.es, \{xavier.gelabert, andreas.hansson\}@huawei.com}
}

\begin{document}
\maketitle

\begin{abstract}
There is a widely held belief that future Radio Access Network (RAN) architectures will be characterized by increased levels of virtualization, whereby base station functionalities, traditionally residing at a single location, will be scattered across different logical entities while being interfaced via high-speed fronthaul (FH) links. For the deployment of such FH links, operators are faced with the challenge of maintaining acceptable radio access performance while at the same time keeping deployment costs low. A common practice is to exploit statistical multiplexing by allowing several cells to utilize the same FH link. As a result, in order to cope with the resulting aggregated traffic, different techniques can be used to reduce the required FH data rates. Herein, we focus on FH compression control strategies for multiple-cell/multiple-user scenarios sharing a common FH link. We propose various methods for sounding reference signal (SRS) handling and analyze different FH-aware modulation data compression and scheduling strategies. Considering a full system setup, including the radio and FH access networks, numerical evaluation is conducted using a 5G NR system-level simulator implemented in ns-3. Simulation results show that, under stringent FH capacity constraints, optimized modulation compression strategies provide significant user-perceived throughput gains over baseline strategies (between 5.2x and 6.9x). On top of them, SRS handling methods achieve additional 2\% to 41\% gains.
\end{abstract}


\IEEEpeerreviewmaketitle

\section{Introduction}
\label{sec:intro}
The design of future Radio Access Networks (RANs) often needs to fulfil the demanding requirements across different and competing axis. While increasing the spectral efficiency has a major impact in the perceived quality of service for the end user, implementing a scalable and low-power solution is often regarded as important for network operators. Furthermore, an efficient use of computing resources via pooling, or the ability to provide cross-layer solutions are also desired features. With this in mind, the Centralized-RAN (C-RAN) architecture paradigm~\cite{Peng16} has emerged and is being considered by 3GPP and Open-RAN (O-RAN) as a key design alternative in next-generation RANs. Among its features, C-RAN advocates for the dissagregation of baseband processing functions between different physical entities which can either be distributed and residing close to the antenna or centralized in a given location. 
Specifically, a base station (a.k.a. a gNB in 3GPP 5G New Radio (NR)) will break up into a Centralized Unit (CU) communicating with at least a Distributed Unit (DU) via the so-called midhaul interface~\cite{TS38401}. In turn, several Radio Units (RUs) will interface towards a DU via the fronthaul (FH)~\cite{oranFH}.
When designing and deploying C-RAN, it is important to consider both capacity constraints and latency requirements of the FH~\cite{larsen:19}. More so, considering the increased bandwidths in 5G NR, in addition to antenna densification, increased modulation orders and enhanced carrier aggregation features~\cite{parkvall:17}. All of these contributing to the increase in the required FH capacity~\cite{lagen:21}.

In general, the dimensioning of the FH may respond to the peak rate requirements given during the planning phase of a specific network technology deployment (e.g. 4G LTE). Nonetheless, under normal network operation, several reasons exist causing the FH to undergo resource under-provisioning at specific moments in time~\cite{7096298}. One example is the ever-growing adoption of new features, new functional splits, or new algorithms requiring increased information exchange between RUs and DUs/CUs. Other examples are, to allow a seamless roll-out of new radio access technologies (e.g., 5G NR), or to facilitate a continuous layout of low-power small cells in specific traffic-demanding areas while at the same time allowing to handle this new data through  the available and pre-dimensioned FH network.
From the equipment vendor's viewpoint, some interest may be rooted in offering a set of new features requiring minimal disturbance on the existing FH network, where only software updates would be necessary to upgrade firmware, algorithms, etc. On the contrary, a dimensioning change in the FH network may be seen as a costly measure, that is, by exchanging optical interface adapters, network switches and, at worst, optical fiber layouts themselves (e.g., switching from single-mode to multi-mode). All the above provides a good motivation to consider the case where the FH can run into a capacity under-provisioning problem.

To lessen the demand in FH capacity and address the above-mentioned FH under-provisioning problem, FH compression schemes become essential~\cite{Peng16}. 
Briefly, FH compression involves the partial reduction, or total removal, of information sent over the FH. FH compression methods have received a wide attention from both the information theory and signal processing communities, in particular after the emergence of C-RAN architectures in 4G LTE. Recently, more practical schemes have been defined in 3GPP 5G NR and O-RAN~\cite{oranFH}. Among the envisioned techniques, \textit{modulation compression} is highlighted given its ability to reduce the required FH capacity owing to constraining the modulation order. This can be effectively achieved  with no degradation of the transmitted samples sent over the FH network and with reduced algorithmic complexity~\cite{lagen:21}. 

Besides FH compression, aimed at jointly reducing the deployment and operational costs along with fostering deployment scalability, the ability to multiplex and aggregate data from multiple cells over a single shared FH interface is of great interest for mobile network operators~\cite{oranDocomo}. In this case, the same fixed FH link is shared by multiple cells (i.e., data from/towards several RUs are multiplexed by using a layer-2 switch over a single FH link towards/from their respective DUs), as shown in Fig.~\ref{fig:depl}. These scenarios present multiple technical challenges arising from the shared FH use by data originating or terminating from/to different cells. Essentially, effective FH compression control schemes have to especially consider and exploit the fact that many cells share a fixed capacity full duplex FH link, and should enhance the multiplexing gain by using clever combinations of different compression techniques, by providing efficient methods to control the use of FH resources while minimizing air-interface performance degradation. Consequently, an evaluation approach relying on end-to-end system-level simulations in a multi-cell environment is carried out in this work.
In~\cite{lagen:21tmc}, baseline along with improved FH-aware packet scheduling methods for dynamic modulation compression were derived. Therein, both the scheduling and modulation compression decisions were dynamically adjusted according to the monitored FH capacity. 

When considering the full system, even if we focus on downlink data transmission, the big part of the data bulk sent through the FH interface comes from the Physical Downlink Shared Channel (PDSCH) (used to send data) and the uplink Sounding Reference Signals (SRSs) (used to estimate the channel). Modulation compression helps reduce the PDSCH bulk part in the FH downlink link. However, when using SRS-based channel estimation for beamforming/precoding design, as standardized for Time Division Duplex (TDD) 5G systems, the FH uplink utilization can be reduced by selectively compressing/removing unnecessary SRSs, which may further impact the beamforming/precoding design in the downlink. Here, \textit{SRS handling} techniques can alleviate the FH uplink load, by properly handling the allocation of SRS signals through the full duplex FH interface. 
In this paper, differently from~\cite{lagen:21} where disaggregated architectures and FH compression methods were reviewed, and from~\cite{lagen:21tmc} where dynamic modulation compression methods were proposed to reduce only the PDSCH bulk, we provide a summary and a new vision of integral solutions for dynamic FH compression control in shared FH architectures. In particular, we review methods for FH-aware downlink data scheduling and we provide a new study for SRS handling methods in the uplink. Finally, the numerical evaluation of the aforementioned schemes are carried out using a seasoned dynamic system-level simulator developed in ns-3~\cite{5glena}. 

The rest of the paper is structured as detailed hereafter. Sec.~\ref{sec:design} discusses the system model and overviews FH compression methods, with special attention to SRS handling methods for uplink SRSs. Sec.~\ref{sec:ns3} describes the simulation scenario and assesses the end-to-end performance. Finally, Sec.~\ref{sec:fut} highlights future research lines and Sec.~\ref{sec:conc} concludes the paper.

\section{Fronthaul Compression Control}
\label{sec:design}
This section introduces the system model and discusses solutions for shared FH architectures integrating FH-aware scheduling methods and SRS handling methods, to reduce the FH downlink and FH uplink loads, respectively.

\subsection{System model}
\begin{figure}[!t]
  \centering
  \includegraphics[width=0.98\linewidth]{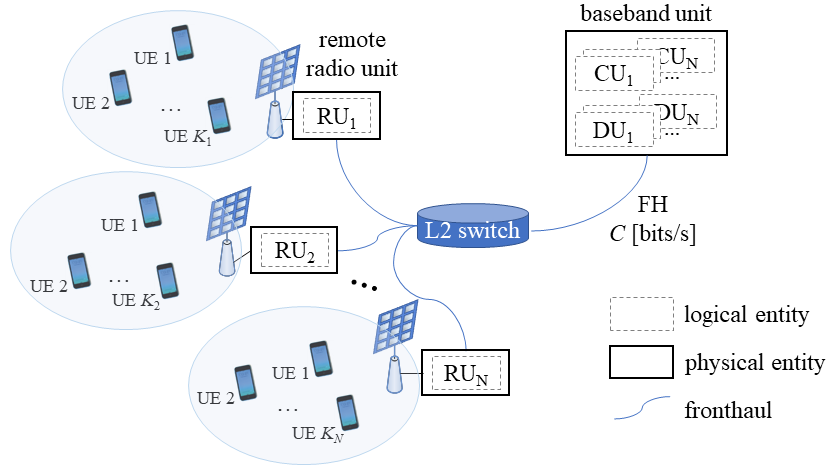}
  \caption{Deployment scenario. Multiple cells and multiple UEs per cell. Hybrid C-RAN architecture with multiple CUs, DUs, and RUs. Star FH topology with a shared full duplex FH interface of limited capacity in downlink and uplink.  } 
  \label{fig:depl}
\end{figure}

We consider herein a multiple-cell/multiple-user (UE) cellular deployment following a C-RAN architecture and consisting of co-located CUs/DUs and geographically distributed RUs. We consider the PDCP-RLC split (a.k.a. Option 2) for the CU-DU~\cite{TS38401}, and the intra-PHY split (Option 7.2x) for the DU-RU~\cite{oranFH}, as per 3GPP and O-RAN specifications respectively. Regarding the CU/DU/RU deployment, we follow the so-called Scenario B, as highlighted by O-RAN~\cite{oranWP}. In this case, 
CUs and DUs (for all cells) are located together in a centralized location (edge or regional cloud), whereas RUs are scattered at operator-owned cell sites. Consequently, the centralized entity implements the high-PHY, MAC, and above processing for all cells, while the RU of each site executes the low-PHY and RF processing of every cell~\cite{TR38801}. 
DUs are interconnected with the RUs through a low layer full-duplex FH interface of limited capacity in downlink and uplink directions, according to a star FH topology~\cite{oranFH}. This way, RUs share the same full duplex FH link. The deployment scenario is illustrated in Fig.~\ref{fig:depl}. 

We focus on FH compression control for downlink data transmission in multi-cell TDD systems with a shared full duplex FH interface. In particular, we 1) analyze FH-aware scheduling methods based on modulation compression to meet the available FH downlink capacity and 2) propose SRS handling methods to meet the available FH uplink capacity. 

\subsection{Downlink FH-aware scheduling methods for PDSCH}\label{DL}
In typical cellular systems, the downlink Modulation and Coding Scheme (MCS) for each UE is determined based on the reported Channel Quality Indicator (CQI). Given the MCSs and the amount of data in the RLC buffers, the MAC scheduler decides the number of resource blocks (RBs) allocated to each UE by following specific scheduling rules (e.g., proportional fair). To meet the shared FH downlink capacity constraint, several FH-aware scheduling methods have been proposed in the literature. Two baseline options are 1) dropping packets at the PHY layer and 2) postponing scheduling decisions at the MAC layer~\cite{lagen:21tmc}. Another option, is to limit the MCS per cell~\cite{lagen:21}, using modulation compression. To enhance these solutions, centralized optimized methods have been proposed in~\cite{lagen:21tmc}, in which the resource allocation (i.e., number of RBs) and modulation compression (i.e., MCS) of each UE are dynamically set. These methods are reviewed in what follows.	

\subsubsection{Drop packets at high-PHY}\label{drop} 
Assuming typical MCS selection and RB assignment at the MAC scheduler, packet dropping can be implemented at the high-PHY layer in the DUs. In this case, a centralized logic determines to drop those MAC PDUs (including new data and/or HARQ retransmissions and their associated control, across all cells) that cannot fit in the available shared FH downlink capacity, cf.~\cite{lagen:21tmc}.

\begin{figure*}[!t]
  \centering
  \includegraphics[width=0.7\linewidth]{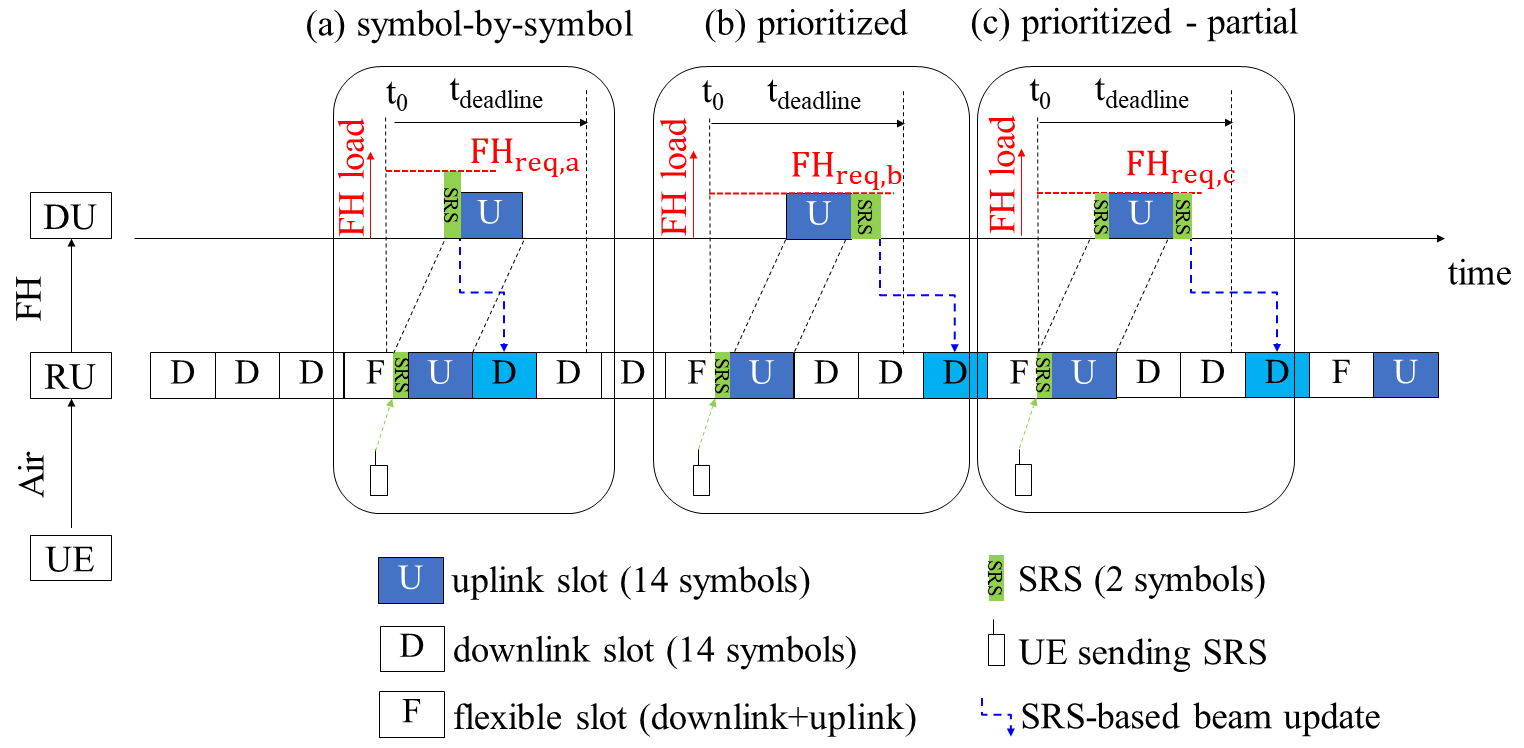}
  \caption{SRS transmission through the full-duplex FH interface, when using a TDD pattern of [D D D F U] in the air interface  (i.e., three downlink slots, followed by a flexible slot, and one uplink slot) and SRSs being sent in the F slots. (a) symbol-by-symbol transmission, (b) prioritized FH transmission, (c) prioritized FH transmission with partial SRS transmissions.} 
  \label{fig:srs}
\end{figure*}

\subsubsection{Postpone scheduling decisions at MAC}\label{postpone}
Assuming typical MCS selection and RB assignment at the MAC scheduler, discarding/dropping of the MAC scheduling decisions can be executed at the MAC layer in the DUs. This way, data is not dropped but its transmission (or retransmission) is postponed. In this case, a centralized logic determines to drop/discard those scheduling decisions (related to new data and/or HARQ retransmissions, across all cells) that cannot fit in the available shared FH downlink capacity, for which its associated transmission is postponed~\cite{lagen:21tmc}.

\subsubsection{MCS limits at MAC}
By exploiting semi-static modulation compression~\cite{lagen:21}, the system can limit the maximum MCS (and so, the maximum modulation order) that is allowed per cell according to the available shared FH downlink capacity. Note that per-cell MCS limits can be combined with dynamic methods, like drop and postpone strategies. In particular, their joint operation may result in less packet drops and scheduling decisions postpones because of the inherent FH load reduction when using lower MCSs.

\subsubsection{Resource allocation and MCS optimization at MAC} \label{sec:opt}
By using dynamic modulation compression, a
centralized control entity can manage the MAC schedulers of all the cells (placed in the collocated DUs in Fig.~\ref{fig:depl}) and determine the most appropriate MAC scheduling decisions (including scheduling of users, MCS assignment, and resource allocation) across all cells, dynamically, so that the shared FH downlink capacity is properly exploited and certain quality-of-service per user is satisfied.
In particular, two solutions are derived in~\cite{lagen:21tmc}, to optimize the RB allocation and the modulation compression applied to each UE of each cell, for every time instant. 

\subsection{Uplink FH-aware handling methods for SRSs}\label{srs}
In TDD systems, SRSs can be used for beam management in downlink and uplink, owing to beam and channel reciprocity. In particular, SRS receptions at the base station are typically used to estimate the channel and then determine the beamforming/precoding.
To get an accurate acquisition of the SRS signal, the bulk needed to send SRSs through the FH uplink interface (RU-to-DU) can be very large, since its size depends on the number of antennas used for channel estimation and the number of resource elements carrying SRS samples.

A key observation is that SRS signals, differently from downlink/uplink data in PDSCH/PUSCH, are non-delay sensitive and do not need to be transmitted through the FH as soon as they arrive at the RU. Basically, we can exploit the fact that the FH interface is full duplex, while the air interface is half duplex, and so, by leveraging on the TDD pattern, SRSs can be sent through the FH uplink interface when there is a downlink slot. 

\begin{figure*}[!t]
  \centering
  \includegraphics[width=0.88\linewidth]{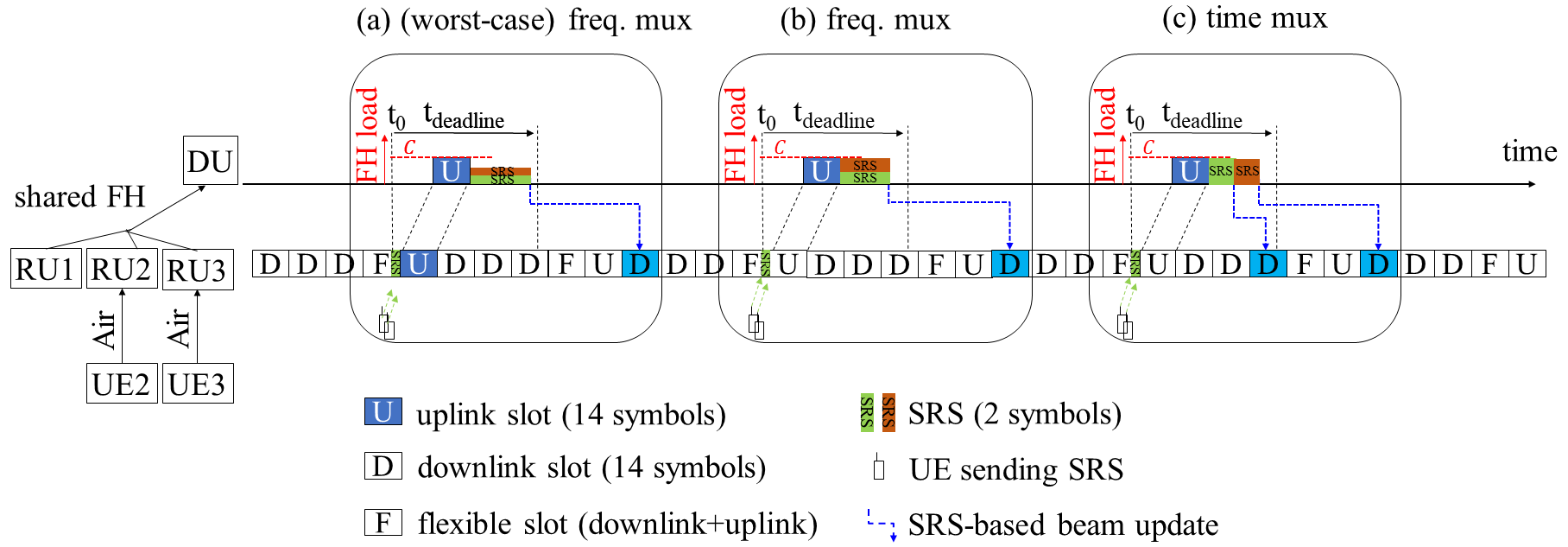}
  \caption{SRS handling methods for SRS bulk transmission through the FH. As an example, the FH is shared among three RUs, being two active at the F slot. (a) (worst-case) fixed frequency multiplexing, (b) dynamic frequency multiplexing, (c) dynamic time multiplexing. }
  \label{fig:srs2}
\end{figure*}

Fig.~\ref{fig:srs} illustrates examples of the impact on the peak FH bandwidth requirements of SRS transmissions, assuming that for a given SRS signal, a certain amount of bits needs to be processed at the DU before some target deadline. There are two main examples of how to meet the latency constraint:
\begin{itemize}
    \item Symbol-by-symbol FH transmission (Fig.~\ref{fig:srs}a): After each symbol is received at the RU, it is immediately transmitted over the FH. The required FH capacity is then given by the maximum between the SRS and PUSCH bulk requirements. This option causes high load peaks on the FH, and makes that the dimensioning of the required FH capacity is determined by SRS bulks, which require more samples than PUSCH, especially for multi-antenna and wide-bandwidth systems.
    \item Prioritized FH transmission (delayed transmission) (Fig.~\ref{fig:srs}b-c): A certain delay is allowed when conveying SRSs over the FH. For example, in Fig.~\ref{fig:srs}b, PUSCH transmission is prioritized and SRS samples are buffered at the RU and transmitted later over a time that may be longer than 2 symbols but still allows the SRSs to be processed before its deadline. A further enhancement is to partially transmit the SRSs, before and after the full PUSCH bulk, as shown in Fig.~\ref{fig:srs}c. In both cases, we need buffering at the RU for SRSs and the FH uplink capacity requirement is determined by the PUSCH bulk, and not by the more demanding SRS bulk. 
\end{itemize}
Prioritized FH transmissions reduce the FH uplink capacity requirement at the cost of a different delay on the reception of the full SRS signal at the DU. This consequently implies a delay on the beamforming/precoding update (illustrated by the light blue downlink slot in Fig.~\ref{fig:srs}), which may affect the downlink end-to-end performance and should be properly evaluated through system-level simulations.

In case of multiple RUs sharing the same FH interface, and potentially overlapping SRS transmissions (e.g., because different cells/RUs use the same TDD pattern), SRS handling methods needs to be designed. Two options appear: time multiplexing or frequency multiplexing.
In the frequency multiplexing option, all the SRS bulks experience the same FH delay, as shown in Fig.~\ref{fig:srs2}a-b. We can use a worst case partition of the available FH bandwidth among the multiple RUs that share the FH (as shown in Fig.~\ref{fig:srs2}a, for the case of 3 RUs sharing the FH interface, where though only 2 RUs send SRSs at the same time). In this case, the FH capacity may not be fully exploited in the uplink direction. Otherwise, we can adopt a dynamic FH bandwidth allocation to the RUs that have data to be sent on a particular slot (as shown in Fig.~\ref{fig:srs2}b), where the total delay can be reduced, compared to the hard bandwidth distribution option. On the other hand, through the time multiplexing option, dynamicity is naturally achieved. In this case, SRS bulks are sent sequentially (see Fig.~\ref{fig:srs2}c), so that some of them are received more quickly at the DU/CU for its processing. This option allows the beam update to be done sooner, compared to the frequency multiplexing options. Here, SRS priority handling methods are needed, to decide the order/priority to send the UEs’ SRS bulks.

\section{End-to-End Simulation}
\label{sec:ns3}
For the evaluation, ns-3 5G-LENA system-level simulator is used~\cite{5glena}. We extended the 5G-LENA simulator with a new centralized intelligence that controls the MAC/PHY operations of all the DUs and implements the proposed FH-aware scheduling procedures and SRS handling methods.

\subsection{Scenario}
We consider an hexagonal site deployment with three sites, according to a Urban Micro scenario. Each site is composed of 3 cells and 3 uniform planar antenna arrays, covering 120º in azimuth each. Frequency reuse 1 is assumed. 
The rest of deployment and network parameters are detailed next:
\begin{itemize}
    \item Number of cells (RUs): 9
    \item Number of UEs per cell: 10
    \item Inter-site distance: 200 m
    \item RU antenna height: 10 m
    \item RU transmit power: 30 dBm
    \item RU antenna: $5\times2$ directional elements 
    \item UE antenna height: 1.5 m
    \item UE antenna: 1 isotropic element
    \item Carrier frequency: 2 GHz
    \item Bandwidth: 100 MHz
    \item Numerology: 1 (30 KHz subcarrier spacing)
    \item RB overhead: 0.04
    \item Duplexing mode: TDD, with  pattern [D D D F U].
    \item SRS: in F slot, spanning over 1 OFDM symbol. Two SRS periodicities: 
    \begin{itemize}
        \item 50 ms (\textit{SRS config1})
        \item 25 ms (\textit{SRS config2})
    \end{itemize} 
    \item MAC scheduler: Round Robin
    \item MCS Table: 2 (up to 256QAM)
    \item channel update period: 40 ms
    \item HARQ: Incremental Redundancy, 20 HARQ processes 
    \item RLC: Unacknowledged Mode
    \item Transport protocol: UDP
    \item Traffic: File Transport Protocol (FTP) Model 1~\cite[Sec. A.2.1.3.1]{TR36814}, YouTube video characterization~\cite{samsungppt}:
    \begin{itemize}
        \item file size: 50 KBytes
        \item file generation rate: $50$ files/second
    \end{itemize}
    \item FH: start topology, shared full-duplex FH link of 0.5 Gbps capacity in each direction (downlink/uplink)\footnote{The FH capacity has been derived based on the peak FH throughput computation. Considering 9 cells, all RBs/symbols used with maximum MCS, and multiplexing gain of 0.5, the total peak FH throughput results 3.6 Gbps. Thus, we select 0.5 Gbps value, to increase the probability that the system is constrained by the FH.}
    \item Simulation duration: 10 s
\end{itemize}

As key performance indicator we consider the ``user-perceived throughput'' (UPT), measured at the IP layer. The UPT corresponds to the fraction between the received bytes per file and the time period needed to complete the file transfer. 

\subsection{Results}
In the end-to-end evaluation, we assess the impact of using different FH compression control methods. For downlink data, we consider the following FH-aware scheduling methods (described in Sec.~\ref{DL}):
\begin{itemize}
    \item Drop: High-PHY drop of MAC PDUs.
    \item Postpone: Discard MAC scheduling decisions. 
    \item RB: RB assignment optimization per active UE. 
    \item MCS: MCS optimization per active UE. 
\end{itemize}
We evaluate each strategy in combination with the following SRS handling methods (detailed in Sec.~\ref{srs}):
\begin{itemize}
    \item fixedDelay: the FH uplink bandwidth is equally distributed among all the cells that share the FH.
    \item dynDelay freqMux: the FH uplink bandwidth is equally distributed among active cells (delay is time dependent).
    \item dynDelay timeMux: the FH uplink bandwidth is fully used by each SRS bulk (delay is time and cell dependent).
\end{itemize}

\begin{figure}[!t]
  \centering
  \includegraphics[width=1\linewidth]{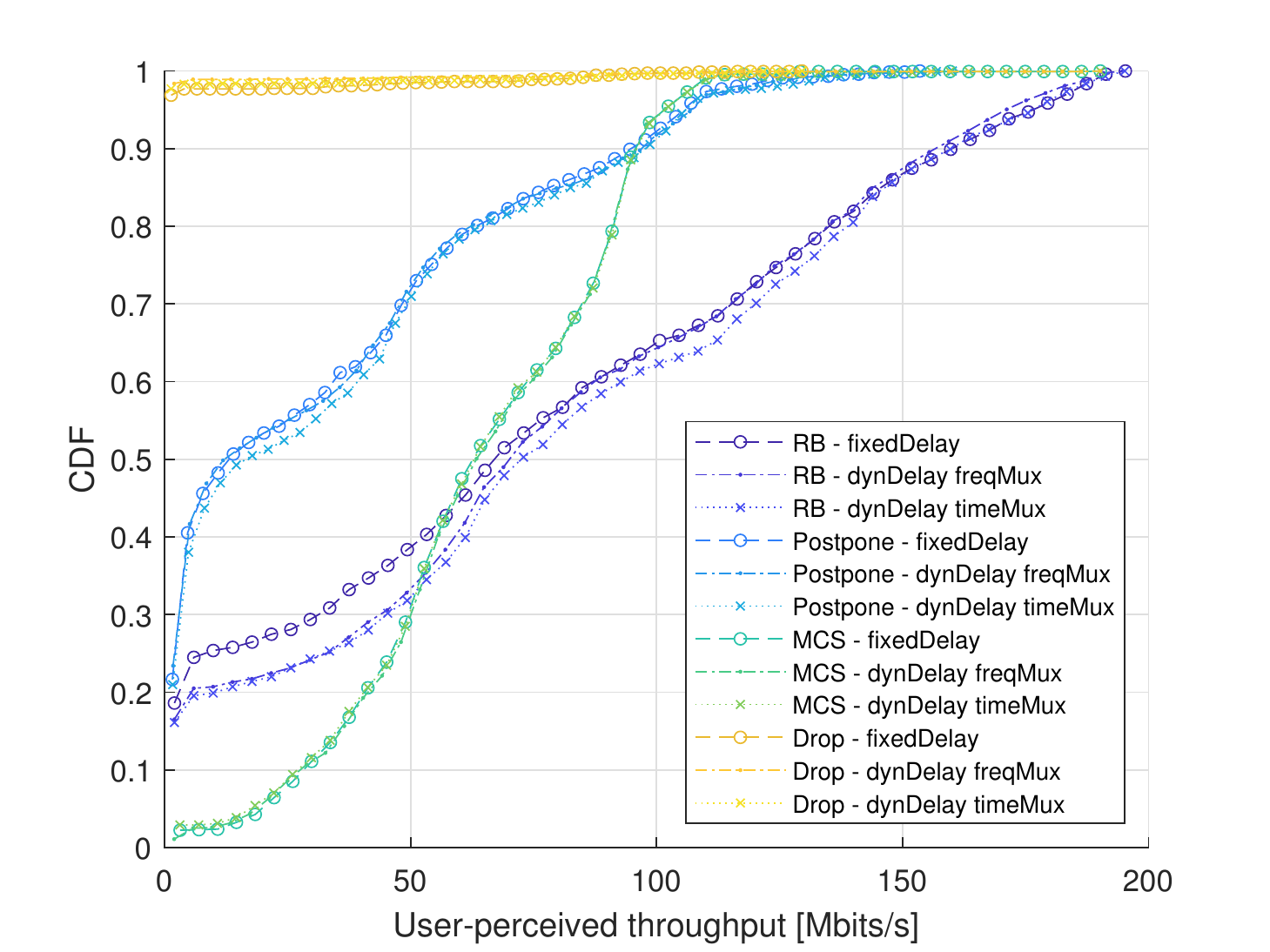}
  \caption{UPT CDF (Mbps), for different scheduling and SRS handling methods. SRS config1.} 
  \label{fig:config1}
\end{figure}

\begin{figure}[!t]
  \centering
  \includegraphics[width=1\linewidth]{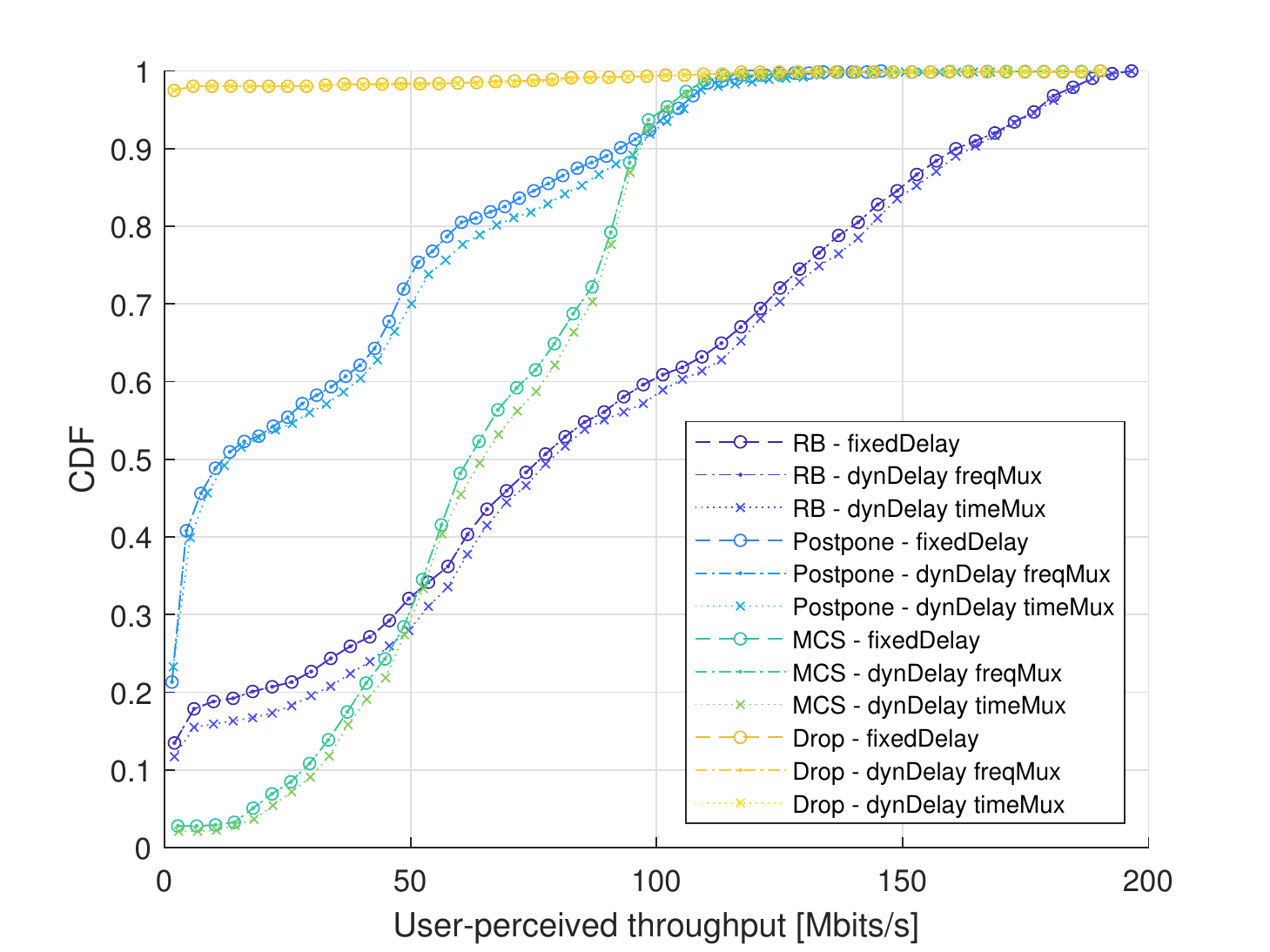} 
  \caption{UPT CDF (Mbps), for different scheduling and SRS handling methods. SRS config2.} 
  \label{fig:config2}
\end{figure}

Fig.~\ref{fig:config1} shows the Cumulative Density Function (CDF) of the UPT (in Mbps) when using SRS config1, for different scheduling strategies and SRS handling methods. Fig.~\ref{fig:config2} displays the same for SRS config2. 

Regarding FH-aware scheduling methods, the Drop technique exhibits very bad end-to-end performance. This is because it performs the scheduling as usual and, in many cases, the scheduled data allocations cannot fit in the available FH bandwidth, implying frequent data drops. The performance is improved with Postpone strategy, which postpones the scheduled allocations for moments in which FH capacity is available, leading occasionally to increased delay. In the considered scenario, characterized by a tight FH capacity, a clear advantage is observed with RB and MCS optimization strategies over baseline strategies (Drop and Postpone) (see Fig.~\ref{fig:config1} and Fig.~\ref{fig:config2}). The benefits are significant in all percentiles of the UPT, ranging from 5.2x to 6.9x in the mean UPT over the Postpone strategy. Indeed, the MCS strategy outperforms the RB strategy in the 5\%tile UPT because, by reducing the modulation order, higher robustness is achieved against propagation/interference variations. Conversely, RB strategy outperforms MCS in the 95\%tile UPT, because it provides a more efficient RB distribution, due to a fine grained control mechanism, thus enabling a larger amount of data being served. 

Regarding the SRS handling methods, we observe that the impact of using beams that are not well adjusted to the channel is appreciable in the RB optimization strategy, while for other strategies the impact is reduced (see Fig.~\ref{fig:config1} and Fig.~\ref{fig:config2}). This is because Drop/Postpone strategies are dropping/postponing many packets and so the performance is dominated by the losses; while the MCS strategy is more robust to signal-to-interference-plus-noise ratio (SINR) degradation. 

The impact of the different SRS handling methods on the UPT depends on the SRS configuration. Specifically, in SRS config1 (see Fig.~\ref{fig:config1}), both the frequency and time multiplexing with dynamic adaptation improve the UPT performance of the worst-case frequency multiplexing, and both provide similar performance. This is because with the considered frame pattern and SRS periodicity (50 ms), there are 20 available slots for SRSs within the SRS periodicity and the deployment considers 10 UEs per cell. Therefore, not all the cells are active in each SRS opportunity. In this way, the dynamic freqMux also allows reducing beam update delays compared to the worst-case freqMux option, and gets similar
UPT performance as the timeMux option, which has lower delays, because
SRS bulks are sent one after the other.
Under SRS config2, we have considered the same scenario but with an SRS periodicity of 25 ms, which results in 10 opportunities for SRSs. All
SRS opportunities are then used by one of the UEs to send SRSs. In this case, as expected and shown in Fig.~\ref{fig:config2}, the two frequency multiplexing options (with fixed or time-dependent delays) achieve the same end-to-end performance, because the two options are equivalent. Interestingly, the time multiplexing option outperforms both frequency multiplexing options, since major part of the SRS bulks experience lower delay updates. In summary, dynamic SRS handling methods achieve mean UPT gains ranging from 2\% to 41\%.

Finally, the presented results have allowed interesting observations regarding the behaviours of FH-scheduling and SRS-handling methods, as a function of the type of served UEs (i.e., cell-edge, cell-middle, or cell-center). A summary of main conclusions is shown in Table~\ref{tab:comp}. For cell-edge UEs, the best option is to use MCS strategy and the SRS handling method does not impact the performance. For cell-middle and cell-center UEs, the best option is to use RB strategy in downlink. In uplink, the cell-center UEs are not affected by the SRS handling method, while for cell-middle UEs the recommended option is the time multiplexing. 

\begin{table}[!t]
\small
\centering
\caption{Best FH-aware scheduling strategy and best SRS handling method, depending on the UE position within a cell.}
\label{tab:comp}
\begin{tabular}{m{2.8cm}||m{1.2cm}|m{1.5cm}|m{1.4cm}}
 & cell-edge & cell-middle & cell-center \\
 \hline
FH-aware scheduling & MCS  & RB & RB  \\
 \hline
 SRS handling & all  & dynDelay timeMux & all  \\
\hline
\end{tabular}
\end{table}

\section{Future Research Directions}
\label{sec:fut}
Based on the presented study, analysis, and obtained simulation results, we envision the following research lines:
\begin{itemize}
    \item SRS priority handling methods:
    The results in Sec.~\ref{sec:ns3} have shown that users in different conditions (cell-edge/middle/center) are differently affected by the delay updates of the beams. Accordingly, in case a time-multiplexing option is adopted, clever SRS priority handling methods for the SRS bulks have to be defined. Under a shared FH capacity,  SRS bulks associated to specific UEs could be prioritized to reduce their delays. For example, when using RB strategy, cell-edge UEs could be prioritized, because they are more affected by a delay increase in the beam update. Instead, when using a combined MCS/RB strategy, cell-middle UEs could be prioritized. A control entity at the FH interface could implement the SRS priority handling method, by knowing the SINR associated to each SRS bulk, to distinguish among cell-center/middle/edge UEs.
    \item SRS control methods: The control entity could also keep track of the actual FH delay for each SRS bulk, leading to SRS control methods. For example, if the buffering delay surpasses the channel coherence time, then such an SRS bulk could be dropped from the buffer of packets to be sent through the FH, because when the DU would receive it, the measurement would be already outdated. This would leave FH capacity available for other transmissions.
    \item SRS priority handling methods with partial transmissions and partial beam updates: Partial transmissions of the SRS blocks, as shown in Fig.~\ref{fig:srs}c, constitute an improvement for shared FH interfaces. In this case, part of the SRS bulks of different UEs can be sent earlier, so that the DU/CU can do a first channel estimation and beamforming update with partial information (e.g.., half (or part) of the SRS samples). Later, once the full SRS information is sent through the FH, a second beamforming update can be implemented based on complete information.
     \item Uplink data FH compression: In the present study, we have focused on compression of FH information related to downlink data transmissions, for which the downlink data (in downlink) and SRS (in uplink) constitute the bulk FH part. Future studies could include compression of uplink data in PUSCH.
    \item Joint flexible splits and FH compression: There has been a wide interest on flexible functional split selection recently. However, the interaction between the split selection and the scheduling/resource allocation strategies has been less studied. An interesting area for further research is to analyze joint strategies that optimize the functional split and the FH compression control for shared FH multi-cell scenarios.
\end{itemize}

\section{Conclusions}
\label{sec:conc}
In this paper, we have presented an integral design and a thorough end-to-end evaluation of shared FH scenarios where multiple FH compression control techniques are proposed. In particular: 1) we have analyzed FH-aware scheduling methods, to compress user data that goes through the downlink FH, and 2) we have proposed SRS bulk handling methods, to handle uplink SRS bulks that go through the uplink FH. Then, end-to-end simulations over a 5G-aligned scenario have been presented. In multi-cell scenarios with shared FH link, we have evaluated the impact of four main FH-aware scheduling methods for downlink data compression: Drop, Postpone, RB and MCS, combined with three methods for SRS handling: fixed frequency multiplexing, dynamic frequency multiplexing, and dynamic time multiplexing. Results have shown that, when there is a tight FH capacity, centralized and optimized scheduling strategies (MCS and RB methods) are essential to maintain an acceptable end-to-end user experience. SRS handling methods are shown to affect the RB optimization strategy, for which our results have exhibited that the time-multiplexing option always provides the best performance and improves all the other SRS handling methods, for configurations in which all the SRS opportunities are used to send SRSs. However, when not all the SRS opportunities are used to send SRSs, the dynamic frequency multiplexing can also achieve a similar performance. Interestingly, the degradation in the end-to-end performance depends on the quality/condition of the target UE, and it is more pronounced in the cell-edge/middle users, which get a lower SINR in the downlink as a result of the pathloss degradation and larger errors in the SRS-based channel estimation, for which future research lines have been highlighted.

Based on our findings above, we advocate for the following recommendations. First, operators and vendors should seriously consider the FH under-provisioning problem, whereby a properly dimensioned FH at the planning stage may become under-provisioned over time. Second, considering shared FH link segments is key to exploit multiplexing gains arising from traffic inhomogeneities. Third, when capacity-limited FH problems arise, leveraging on FH compression strategies helps alleviating the problem. We conclude that dynamic compression of data is essential to maintain acceptable user experience, while at the same time noting that compression of reference signals (like, specially, SRSs) plays a relevant role.

\section{Acknowledgment}\small
This work has been partially funded by Huawei Technologies and Spanish MINECO grant TSI-063000-2021-56/57 (6G-BLUR).

\bibliography{references}
\bibliographystyle{ieeetr}

\section*{Biographies}
\begin{IEEEbiographynophoto}{Sandra Lag\'en} holds a 
PhD from UPC (2016). She is a Senior Researcher and Head of the Open Simulations research unit in CTTC.  \\ 
\textbf{Xavier Gelabert} is a senior research engineer at Huawei Technologies Sweden AB. He has 15+ years of experience working across RAN L1, L2 and L3, as well as 3GPP standardisation. \\ 
\textbf{Andreas Hansson}
is a Principal Baseband Software Engineer within Huawei Technologies Sweden AB, working with software architecture, modelling and systemization, with 20 years of experience.\\ 
\textbf{Manuel Requena}
is a Senior Researcher in CTTC and responsible of the EXTREME Testbed of the Services as Networks research unit. \\ 
\textbf{Lorenza Giupponi}
holds a PhD from UPC (2007). She is a Senior Researcher in CTTC, and a member of the CTTC Executive Committee.  
\end{IEEEbiographynophoto}
\end{document}